\documentclass[twocolumn,showpacs,superscriptaddress,amsmath,amssymb]{revtex4}
\usepackage{graphicx}
\usepackage{dcolumn}
\usepackage{bm}
\usepackage{color}
\usepackage{mathrsfs}
\usepackage{ulem}

\begin{document}\large

\title{Cumulants in the 3-dimensional Ising, $O(2)$ and $O(4)$ spin models }

\author{Xue Pan} 
\affiliation{Key Laboratory of Quark and Lepton Physics (MOE) and
Institute of Particle Physics, Central China Normal University, Wuhan 430079, China}
\author{Lizhu Chen} 
\affiliation{Key Laboratory of Quark and Lepton Physics (MOE) and
Institute of Particle Physics, Central China Normal University, Wuhan 430079, China}
\author{X.S. Chen} 
\affiliation{Institute of Theoretical Physics, Chinese Academy of Sciences,
        Beijing 100190, China  }
\author{Yuanfang Wu} 
\affiliation{Key Laboratory of Quark and Lepton Physics (MOE) and
Institute of Particle Physics, Central China Normal University, Wuhan 430079, China}

\setlength\unitlength{1cm}
\begin{picture}(0,0)
\put(12,1.5){\small Submitted to 'Chinese Physics C'}
\end{picture}

\begin{abstract}
Based on the universal properties of a critical point in different systems and that the QCD phase transitions fall into the
same universality classes as the 3-dimensional Ising, $O(2)$ or $O(4)$ spin models,
the critical behavior of cumulants and higher cumulant ratios of the order parameter from the three
kinds of spin models is studied. We found that all higher cumulant ratios change dramatically the sign
near the critical temperature. The qualitative critical behavior of the same order cumulant ratio is consistent in these
three models.

\end{abstract}


\maketitle
\section{Introduction}

One goal of current relativistic heavy ion collision experiments is to locate the QCD critical point. In the idealized
thermodynamic limit, the correlation length $\xi$ would diverge at the critical point~\cite{diverge x}. But the system
size, especially the evolution time of the formed system are finite in relativistic heavy-ion
collisions~\cite{diverge x,slowing x}. $\xi$ may not be fully developed. It's estimated to be at most the value
of 2-3 fm~\cite{slowing x}. It's close to its "natural" value of 1fm. So it's essential to find observable that
is sensitive to the critical point.

Recently, it has proposed that the higher cumulants will be more sensitive to the critical point~\cite{Non-Gaus. flu.}.
On one hand, they are more sensitive to the correlation length. For example, the third cumulant and fourth cumulant of
net-proton in heavy-ion collisions individually diverge with $\xi^{4.5}$ and $\xi^{7}$, which is much faster than the
quadratic cumulant. On the other hand, the higher cumulant can reflect the fine-detail information of the net-proton
distribution. The sign changes of various cumulants are noted in related papers. For example,
the sign of the third order cumulant is discussed in Ref.~\cite{third moments}, the negative fourth order cumulant is
discussed in Ref.~\cite{PLB-Gupta, negative K}, and the sign change of sixth and eighth order cumulants is shown in
Ref.~\cite{six order}. In order to compare the results of theory with experiments, the ratios of higher cumulants to the
second order one is usually used.

As we know, different systems have universal behavior in the vicinity of a critical point. The systems falling to
the same universality class have the same value for the critical exponent. So the results of relatively simple systems,
such as spin models with an $O(N)$-symmetry, play an important role in the analysis of phase transitions in much more
complicated systems.

The critical point terminating the first order phase
transition line in the QCD phase boundary belongs to the same universality class of the 3-dimensional Ising
model~\cite{class 1,class 2, class 3, class 4}. In the chiral limit, if $U_A(1)$ symmetry is broken, the chiral
phase transition of
a 2-flavor QCD theory is argued to belong to the same universality class of the 3-dimensional $O(4)$ spin
model~\cite{class 1}. The critical behavior of the net-baryon number fluctuations is expected to be controlled by
the universal $O(4)$ symmetry group~\cite{symmetry group}.
Because of lattice artifacts in calculations with the staggered fermions, the 2- and (2+1)-flavor chiral phase
transitions may belong to the same universality class with the $O(2)$ spin model~\cite{O2 class 1,O2 class 2,O2 class 3}.
 So the critical behavior of cumulant and higher cumulant ratios
in the 3-dimensional Ising, $O(2)$ and $O(4)$ spin models
are helpful.

In this paper, we first introduce the cumulants in the $O(N)$ spin models and the corresponding relations of the
3-dimensional Ising and $O(4)$ spin models to the QCD.
Then through the Monte Carlo simulations without external magnetic field, we calculate
the cumulants and higher cumulant ratios of the order parameter in the 3-dimensional Ising, $O(2)$ and $O(4)$
spin models. The critical behavior of the order parameter, the susceptibility, the ratios of the third,
fourth and sixth to second order cumulant is presented and discussed, respectively. Meanwhile, the behavior of higher
cumulant ratios in the 3-dimensional Ising, $O(2)$ and $O(4)$ spin models is compared. Finally, the conclusions are drawn.

\section{Fluctuations of order parameter in the $O(N)$ spin models}

The $O(N)$-invariant nonlinear $\sigma$-models($O(N)$ spin model) are defined as,
\begin{equation}\label{Hamiltonian}
\beta \mathcal{H} =-J\sum_{\langle i,j\rangle} \vec{S}_{i}\cdot \vec{S}_{j}-\vec H\cdot\sum_{i} \vec{S}_{i}.
\end{equation}
$\mathcal{H}$ is the Hamiltonian.
$J$ and $\vec H$ are both reduced quantities which already conclude $\beta = 1/T$. $J$ is an interaction energy between
the nearest-neighbor spins $\langle i,j\rangle$. In our simulation, we set $J=\beta$. $\vec H$ is the external magnetic field.
$\vec S_{i}$ is a unit vector of $N$-components at $i$th site of a $d$-dimensional hyper-cubic lattice with a longitudinal
(parallel to the magnetic field $\vec H$) and a transverse component
\begin{equation}\label{spin}
\vec S_i=S_i^\parallel \vec e_H+\vec S_i^\perp,
\end{equation}
where
\begin{equation}\label{unit}
\vec e_H=\vec H/H.
\end{equation}
$H$ is the magnitude of the external field. The cases $N=1, 2, 4$ and $d=3$ are the 3-dimensional Ising, $O(2)$ and $O(4)$
spin models, respectively.
The energy of a spin configuration is defined as~\cite{scaling fun.}
\begin{equation}\label{energy}
E=-\sum_{<i,j>} \vec{S}_{i}\cdot \vec{S}_{j}.
\end{equation}
The average of the longitudinal spin components is
\begin{equation}\label{longitudinal spin components}
S^\parallel=\frac{1}{V}\sum_{i}S_{i}^\parallel,
\end{equation}
where $V=L^3$ and $L$ is the number of spins in each direction.

Then the partition function is
\begin{equation}\label{partition function}
Z(T,H)=\int\prod_i d^NS_i \delta(\vec S_i^2-1)\exp(-\beta E+HVS^\parallel).
\end{equation}
The (reduced) free energy per unit volume is
\begin{equation}\label{freee energy}
f(T,H)=-\frac{1}V \ln Z.
\end{equation}
The derivatives of the free energy density to $H$ are as following,
\begin{equation}\label{chi_n}
\left.\chi_n=-\frac{\partial^nf}{\partial H^n}\right|_{T}.
\end{equation}
They are related with the cumulants of order parameter.
For instance,
\begin{equation}\label{cumulants}
\begin{split}
 & \chi_1={\langle S^\parallel\rangle}, \\& \chi_2=V{\langle \delta {S^\parallel}^2 \rangle},\\& \chi_3=V^2\langle \delta {S^\parallel}^3 \rangle,\\& \chi_4={V^3(\langle \delta {S^\parallel}^4 \rangle-3\langle \delta {S^\parallel}^2 \rangle^2)},\\& \chi_6={V^5(\langle \delta {S^\parallel}^6 \rangle-10\langle \delta {S^\parallel}^3 \rangle^2+30\langle \delta {S^\parallel}^2 \rangle^3-15\langle \delta {S^\parallel}^4\langle \delta {S^\parallel}^2\rangle)}.
\end{split}
\end{equation}
Where
\begin{equation}\label{delta S}
\delta {S^\parallel}=S^\parallel-\langle S^\parallel\rangle,
\end{equation}
$\chi_1$ and $\chi_2$ are respectively the magnetization(order parameter) $M$ and longitudinal susceptibility $\chi_L$.
Owing to the spatial rotation symmetry of the $O(N)$ groups, the mean value of the order parameter is always zero
without external magnetic field. In this case, the order parameter definition should be resorted~\cite{order parameter}, such as
 \begin{equation}\label{order parameter}
M=\langle|\frac{1}{V}\sum_i{\vec S_i}|\rangle.
\end{equation}

The scaling form of the critical part of the free energy in the second order phase transition can be write as
\begin{equation}\label{scaling O1}
f_s(t,h)=l^{-3}f_s(l^{y_t}t,l^{y_h}h),
\end{equation}
where $t$ is the normalized reduced temperature and $h$ is the normalized reduced magnetic field
\begin{equation}\label{parameters O1}
t={(T-T_c)}/T_0,~~ h=H/H_0.
\end{equation}
$T_0$ and $H_0$ are the normalized parameters. $y_t$ and $y_h$ are the thermal and magnetic exponents,
respectively.
For purpose of mapping the results
of the 3-dimensional Ising model to that of QCD, the linear ansatz as following
is suggested~\cite{linear ansats 1, linear ansats 2, linear ansats 3}
\begin{equation}\label{linear relations}
t=T-T_{cp}+a(\mu-\mu_{cp}), ~~~h=b(T-T_{cp})+\mu-\mu_{cp}.
\end{equation}
$T_{cp}$ is the temperature and $\mu_{cp}$ is the chemical potential at the QCD critical point.
$a$ and $b$ have to be determined from QCD. We can get that
\begin{equation}\label{derivatives}
\begin{split}
& \partial/\partial\mu=\frac{-b}{1-ab}\frac{\partial}{\partial t}+\frac{1}{1-ab}\frac{\partial}{\partial h},\\&
\partial/\partial T=\frac{1}{1-ab}\frac{\partial}{\partial t}-\frac{a}{1-ab}\frac{\partial}{\partial h}.
\end{split}
\end{equation}
Through the different orders of derivatives of free
energy density to $\mu$, we can get the fluctuations of particle number, which can be measured by experiments.
The exponent $y_h$ is bigger than $y_t$ in the 3-dimensional Ising model.
Their ratio ${y_h}/{y_t}=\gamma+\beta=25/16$~\cite{Ising exponents}.
So $\partial/\partial h$ will be more singular than $\partial/\partial t$.
Although, we don't know the values of $a$ and $b$, the particle number fluctuations are dominated by the derivatives
of the free energy density to $h$,
that's the order parameter fluctuations in the 3-dimensional Ising model.

The scaling form of the critical part of the free energy density in the chiral phase transition may be
\begin{equation}\label{scaling O4}
\frac{f_s(T,\mu_q,h)}{T^4}=Ah^{(1+1/\delta)}f_f(z), ~~~z=t/h^{1/\beta\delta},
\end{equation}
where $\beta$ and $\delta$ are critical exponents from the 3-dimensional $O(4)$ spin model, $f_f(z)$ is the scaling
function of the free energy density and
\begin{equation}\label{parameter O4}
t\equiv\frac{1}{t_0}(\frac{T-T_c}{T_c}+\kappa_\mu(\frac{\mu_{q}}{T})^2),~~~ h\equiv\frac{1}{h_0}\frac{m_q}{T_c}.
\end{equation}
Here $T_c$ is the chiral phase transition temperature. From Eq.~\eqref{parameter O4}, we note that the derivatives of
free energy density to $h$ in the $O(4)$ spin model are equal to the derivatives of free energy density to $m_q$,
which is the fluctuation of order parameter, or the chiral condensate
\begin{equation}\label{chiral condensate}
\langle\bar{\psi}\psi\rangle = -\frac{N_F}{4} \frac{\partial f}{\partial {m_q}}.
\end{equation}
So the critical fluctuations of order parameter from the 3-dimensional $O(4)$  spin model can reflect the chiral
condensate fluctuations in the QCD chiral phase transition.
When it comes to the chiral limit, the quark masses vanish and the chiral symmetry is restored. It corresponds to
that the magnetic field in the 3-dimensional $O(4)$ model is zero.

\section{The critical behavior of higher cumulant ratios in the 3d Ising, $O(2)$ and $O(4)$ spin models}

The Monte Carlo simulations of the 3-dimensional Ising, $O(2)$, and $O(4)$ spin models in a finite system are performed
by the Wolff algorithm with helical boundary conditions~\cite{wolff algorithm}.
We choose sufficient big size for each case which can present the qualitative features well,
that's $L=24$ for the Ising model and $L=20$ for the $O(2)$ and $O(4)$ spin models.
In order to observe and compare the trend of cumulants and their ratios varying with $T/T_c$ in different models,
we divide the cumulants or their ratios by their maximum values and rescale the values of $\chi_2$, $\chi_3/\chi_2$,
$\chi_4/\chi_2$, $\chi_6/\chi_2$ to plot them. Here $T_c$ is the critical temperature of each model. In our calculation,
we use approximate values 4.51~\cite{order parameter}, 2.202~\cite{O24Tc} and 1.068~\cite{O24Tc} for $T_c$ in the 3-dimensional Ising, $O(2)$ and $O(4)$ spin models,
respectively.

\begin{center}
\includegraphics[width=0.23\textwidth ]{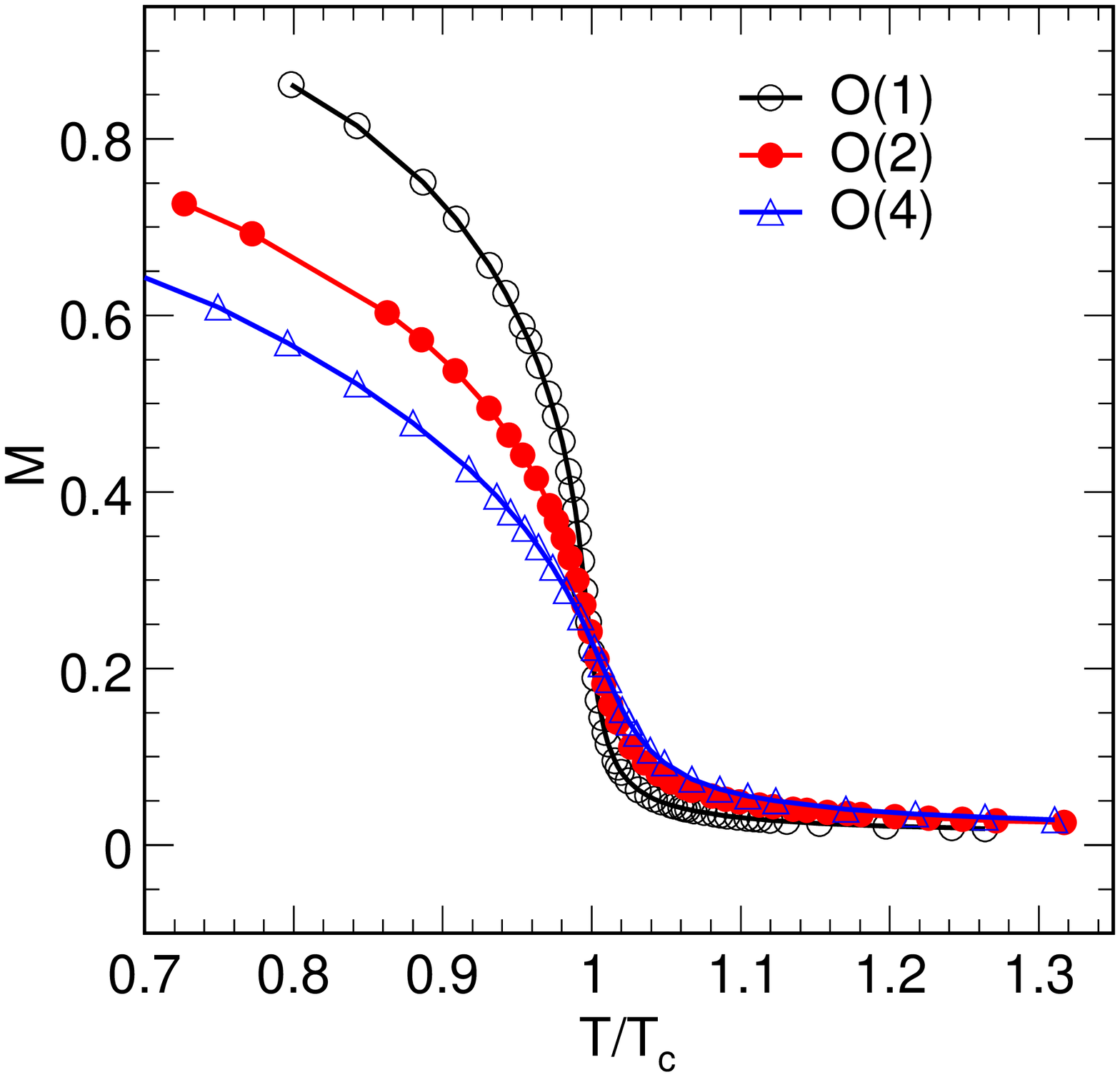}
\includegraphics[width=0.23\textwidth ]{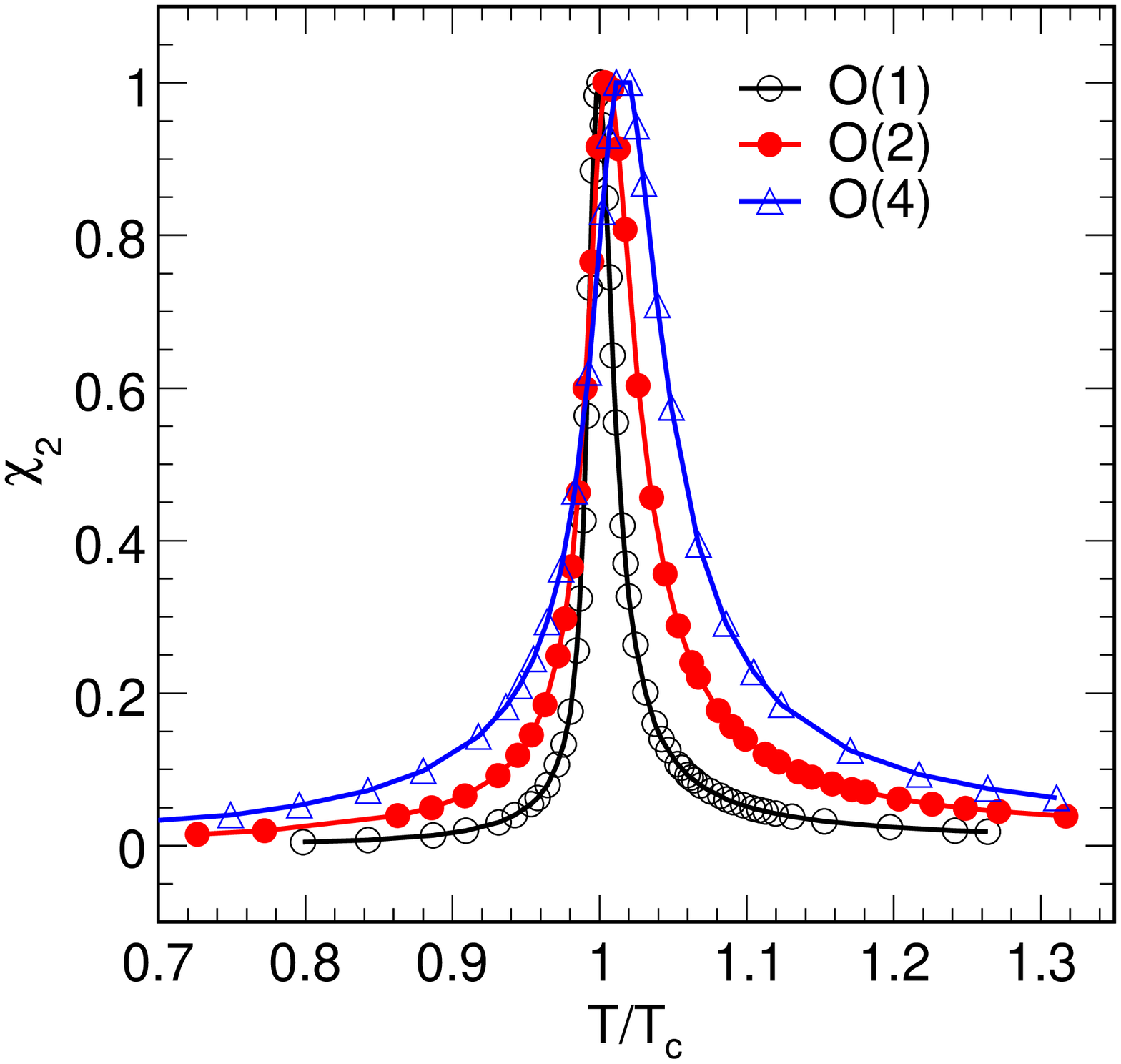}
\end{center}\small

Fig. 1: (Color online) The temperature dependence of the order parameter ($M$) and susceptibility ($\chi_2$) in the vicinity of critical temperature from the 3-dimensional Ising ($O(1)$), $O(2)$ and $O(4)$ spin models.\normalsize

\vspace{15pt}

The order parameter ($M$) and susceptibility ($\chi_2$) from the 3-dimensional Ising, $O(2)$ and $O(4)$ spin models is presented
in Fig.~1. Its behavior in these three spin models is similar. $M$ decreases with the increasing temperature. When the
temperature is much lower than $T_c$, the system is ordered, all of the spins align to the same direction,
the value of $M$ approaches one. When the temperature is much higher than $T_c$, the system is disordered,
the spins point to a direction at random, the value of $M$
approaches zero.
The qualitative behavior of $\chi_2$ in the three models are similar, too. There is a pronounced cusp near $T_c$. The peak
is also observed in the chiral susceptibility in the 2-flavor QCD lattice calculation and the chiral effective model with the Polyakov loop~\cite{suscepbilities in LQCD,suscepbilities in chiral model}.

The ratios of the third ($\chi_3/\chi_2$)
and fourth ($\chi_4/\chi_2$) to the second order cumulant in the 3-dimensional Ising, $O(2)$ and $O(4)$ spin models are shown in Fig.~2.

\begin{center}
\includegraphics[width=0.23\textwidth ]{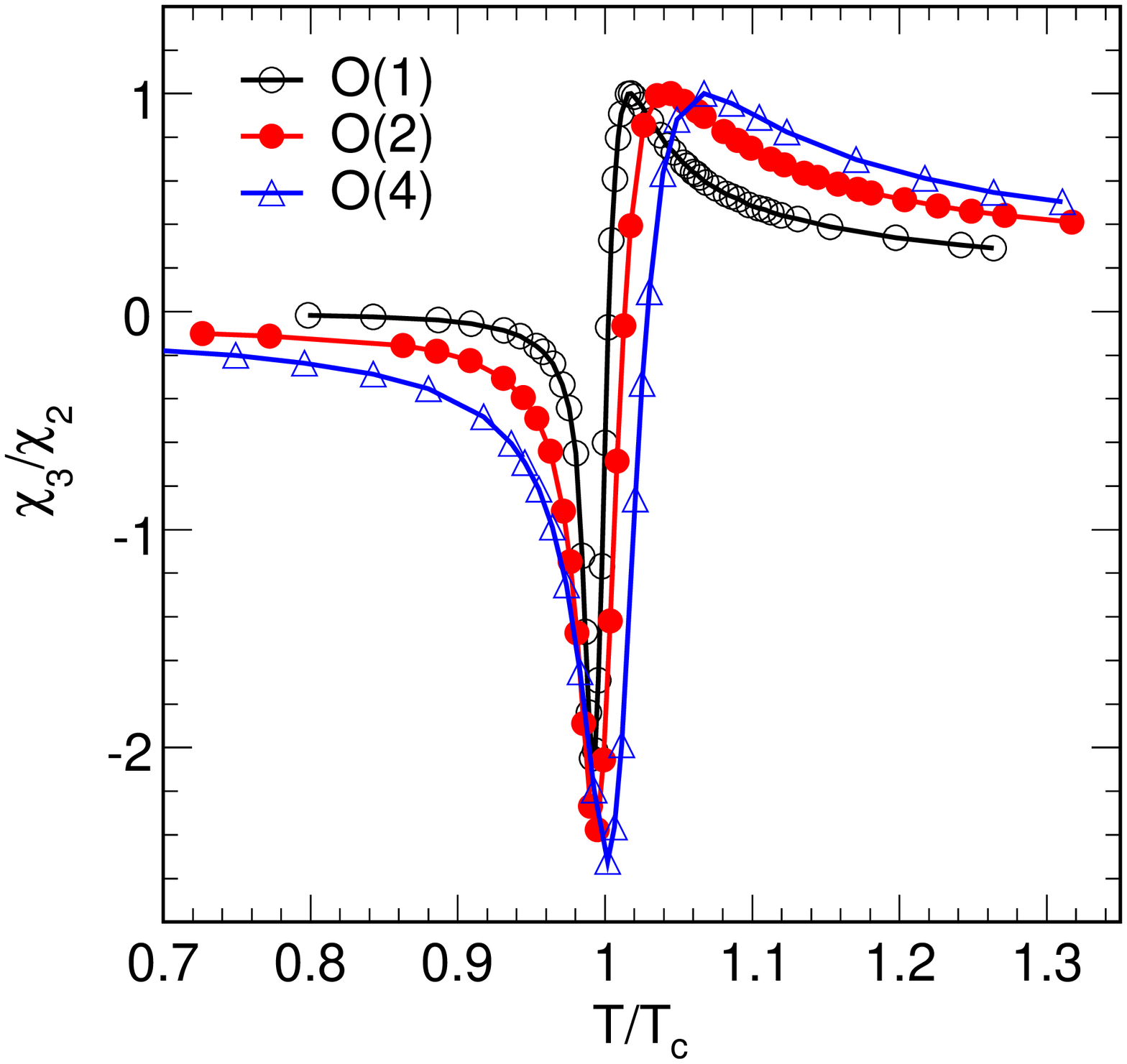}
\includegraphics[width=0.23\textwidth ]{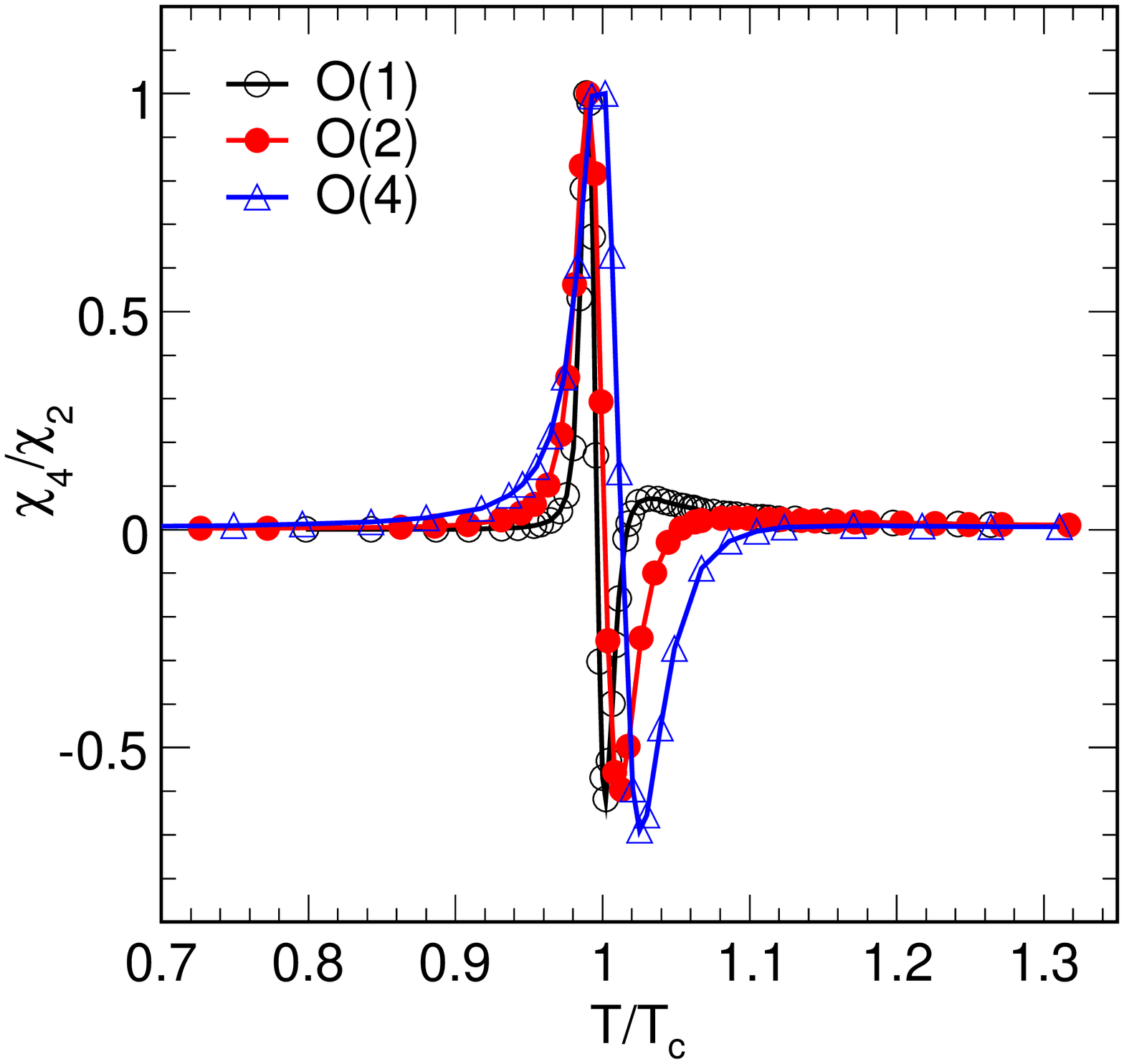}
\end{center}\small

Fig. 2: (Color online) The temperature dependence of $\chi_3/\chi_2$ and $\chi_4/\chi_2$ in the vicinity of
critical temperature from the 3-dimensional Ising ($O(1)$), $O(2)$ and $O(4)$ spin models.\normalsize

\vspace{15pt}

The qualitative behavior of $\chi_3/\chi_2$ is the same in the three models. It changes dramatically the
sign near $T_c$. It is negative when $T<T_c$, and becomes positive when $T>T_c$. The behavior of $\chi_3/\chi_2$ from the 3-dimensional Ising model is consistent with that in Ref.~\cite{cpc}. The third order cumulant reflects the skewness
of the distribution. As the Gaussian distribution, it's symmetrical, then its skewness is zero. If the left tail of a
distribution is longer than the right one, the third order cumulant will be negative. If the right tail is longer, the third
order cumulant will be positive.
The qualitative behavior of $\chi_4/\chi_2$ is the same in the three models. It oscillates greatly with
temperature near $T_c$. They are negative when $T$ approaches $T_c$ from $T>T_c$ side.
This is consistent with the prediction
in~\cite{negative K}, which said the fourth order cumulant will be negative when the system approaches the critical point
from the crossover side. The fourth order cumulant reflects the kurtosis of the distribution. Gaussian distribution is also
the reference. Its kurtosis is zero. If a distribution is less sharp than the Gaussian distribution, its fourth order cumulant
will be negative. If it's more sharp, its fourth order cumulant is positive.

The ratios of sixth to second order cumulant from the 3-dimensional Ising, $O(2)$ and
$O(4)$ spin models are presented in Fig.~3.

\begin{center}
\includegraphics[width=0.35\textwidth ]{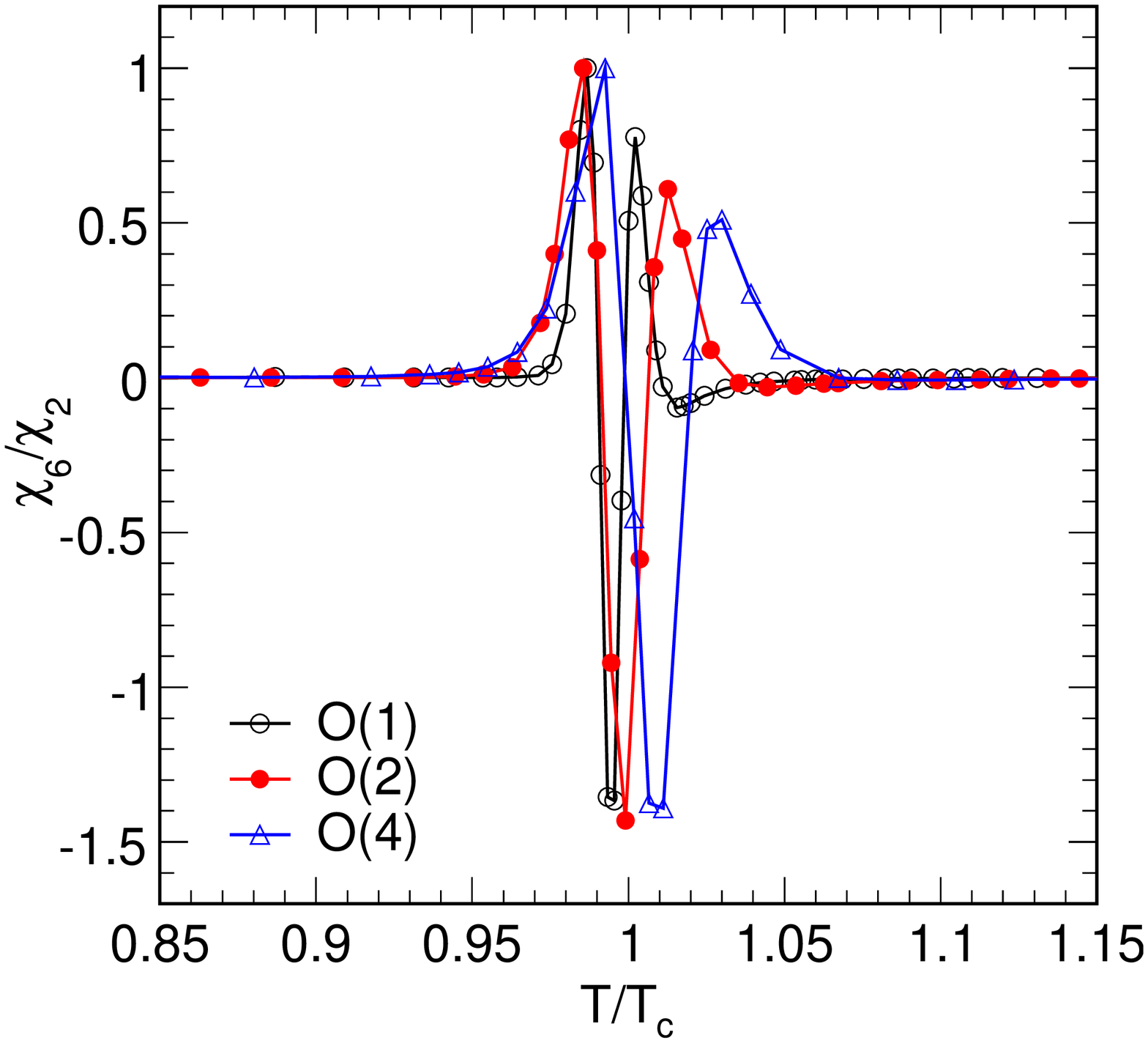}
\end{center}\small

Fig. 3: (Color online)The temperature dependence of $\chi_6/\chi_2$ in the vicinity of
critical temperature from the 3-dimensional Ising ($O(1)$), $O(2)$ and $O(4)$ spin models. \normalsize

\vspace{15pt}

Its generic structure in the three models is similar. It has two positive maximums and a pronounced negative minimum
between them close to the transition region.
Comparing $\chi_3/\chi_2$, $\chi_4/\chi_2$ and $\chi_6/\chi_2$, we found that the higher the order of the
cumulant, the more complicated of the structure and quicker to get equilibrium after leaving the critical point.

From the simulation results, we already know that the behavior of the same order cumulant or higher order cumulant ratio is similar in the three models. Now let's analyze it briefly from the theory. As the order parameter in the spin models, it will approach one and zero in the ordered and disordered phases, respectively, which leads to the similar behavior of the order parameter. In the thermodynamical limit, the susceptibility diverges as $\chi_2 \sim |t|^{-\gamma}$ in the vicinity of the critical point. The value of $\gamma$ is 1.253(4)~\cite{Ising exponents}, 1.3192~\cite{o2 exponents} and 1.4668~\cite{o4 exponents} for the 3-dimensional Ising, $O(2)$ and $O(4)$ spin models. Their values are positive and close to each other. In a finite system, the divergence will be weakened and become a round peak. That's why the behavior of $\chi_2$ is similar. All of the cumulants of the order parameter are derivatives of the free energy density with respect to the external field, the behavior of the first and second order cumulants is similar in these three models, so it's not difficult to understand the similar behavior of the higher order cumulant ratios.

\section{Summary}

In this paper,
the critical behavior of the order parameter, susceptibility, ratios of the third, fourth, sixth order cumulant
to the second one are calculated from the 3-dimensional Ising, $O(2)$ and $O(4)$ spin models
at a given system size without external field. For each order cumulant or higher cumulant
ratios, its qualitative critical behavior in these three models is the same.
The ratios of the third, fourth, and sixth order cumulant to the second one change dramatically near the critical
temperature. They all have sign change. And the higher the order of the
cumulant, the more complicated of the structure. From the 3-dimensional Ising model, we know that the sign changes
of the cumulant ratios of baryon number
fluctuations may predict the critical signals in heavy-ion collisions. For the 3-dimensional $O(2)$ and $O(4)$ spin models,
in order to guide the experiments, the derivatives of free energy density to the temperature in a finite lattice
is on going.

This work was supported in part by the National Natural Science Foundation of China under Grant
No. 10835005, MOE of China under Grant No. IRT0624, B08033 and
Laboratory of Quark and Lepton Physics (MOE) and
Institute of Particle Physics, Central China Normal University, Wuhan 430079, China under Grant No. QLPL201303.

\end{document}